\documentclass[final,5p,times,twocolumn,number]{elsarticle}
\biboptions{sort&compress} 

\usepackage{amssymb}
\usepackage{lipsum}
\usepackage{amsmath}
\usepackage{xcolor}
\usepackage{hyperref}
\hypersetup{colorlinks=true, citecolor=blue, linkcolor=blue}

\begin{document}

\begin{frontmatter}

\title{Stochastic Web Map: Survival probability and escape frequency}

\author[]{$^{1,2}$K. B. Hidalgo-Castro, $^1$J. A. M\'endez-Berm\'udez, $^2$Edson D. Leonel}
\address{$^1$Instituto de F\'isica, Benem\'erita Universidad Aut\'onoma de Puebla, 
Puebla 72570, Mexico\\
$^2$Universidade Estadual Paulista (UNESP) - Departamento de F\'isica,
Av. 24A, 1515, 13506-900, Rio Claro, SP, Brazil}

\begin{abstract}
We study transport and escape in the Stochastic Web Map (SWM), an area-preserving system with phase-space structure controlled by a symmetry parameter $q$ and nonlinearity $K$. By analyzing the survival probability $P_{\text{S}}(n)$ and escape frequency $P_{\text{E}}(\ln n)$, we show that in the chaotic regime escape dynamics is governed by a single time scale $n_{\text{typ}}\propto K^{-2}h^{2}$; here $h$ is the size of the escape horizon. Deviations at large $K$ and small $h$ indicate a breakdown of the quasilinear approximation. Then, upon rescaling the time by $n_{\text{typ}}$, escape statistics becomes universal, independent of $q$. These results demonstrate that escape is controlled by global transport rather than symmetry.
\end{abstract}


\end{frontmatter}




\section{Introduction}
\label{introduction}

The study of transport and escape processes in dynamical systems has proved to be an effective route to uncover global properties of phase space dynamics. A particularly fruitful approach consists in coupling a bounded dynamical system to an asymptotic region by introducing a leak, or hole, through which trajectories escape once they reach a predefined region of phase space \cite{Chirikov267}. In this framework, transport properties can be quantified by monitoring ensembles of trajectories and defining statistical observables such as the survival probability $P_{\text{S}}(n)$, which measures the fraction of trajectories that remain in the system up to time $n$, and the escape time distribution, or escape frequency $P_{\text{E}}(n)$, which characterizes how trajectories leave the interaction region \cite{Chirikov267,Lichtenberg92}.

For classical Hamiltonian systems, the decay of $P_{\text{S}}(n)$ is strongly connected to the underlying phase space structure. In strongly chaotic systems, survival probabilities typically decay exponentially, reflecting fast and nearly homogeneous transport, whereas in mixed phase spaces the presence of sticky regions near invariant structures leads to slower, algebraic decays \cite{Chirikov267,Lichtenberg92}. Most studies of leaking dynamics have focused on systems undergoing the generic Kolmogorov-Arnold-Moser (KAM) transition to chaos, encompassing integrable, mixed, and fully chaotic regimes \cite{Zaslavsky86,Zaslavsky86-2,Zaslavsky87,Chernikov87,Chernirkov87-2}. Within this context, Chirikov's standard map has emerged as a paradigmatic model for exploring transport, diffusion, and escape phenomena \cite{Chernikov88,Lichtenberg89,Lichtenberg89-2}. 

Several signatures of Hamiltonian chaos have been identified through discrete-time mappings, which provide simplified yet powerful representations of nonlinear dynamics \cite{Altmann13}. Among them, Chirikov's standard map has played a central role in the study of the generic transitions to chaos predicted by KAM theory \cite{Livorati12}. Nevertheless, there exist important classes of Hamiltonian systems for which the assumptions of the KAM theorem are not fulfilled, and consequently the standard transition to chaos does not occur. Examples include discontinuous maps and web maps, whose dynamics may exhibit global diffusion mechanisms not constrained by invariant tori \cite{Dettmann12,Bunimovich05}.

Beyond the KAM framework, an important class of systems is characterized by discontinuities or non-smooth phase space structures, often leading to unbounded diffusion and anomalous transport properties \cite{Borgonovi98,Mendez12}. Representative examples include billiard systems with singular geometries, such as the Bunimovich stadium and triangular billiards, as well as discrete-time maps such as the sawtooth map \cite{Borgonovi96,Casati99,Casati00,Prosen01,Casati99-2,Dana89}. Despite the extensive literature on KAM-type systems, leaking dynamics in non-KAM scenarios have only recently attracted systematic attention \cite{Altmann13}.

A closely related and extensively studied paradigm in nonlinear dynamics is the kicked rotor, whose stroboscopic dynamics is described by Chirikov's standard map \cite{Ott08,Chirikov69}. This model has been generalized in several directions to incorporate mechanisms beyond the classical KAM picture, including dissipative, discontinuous, and fractional extensions \cite{Zaslavsky78,Borgonovi98,Tarasov08,Edelman09}. These generalized mappings display rich transport behaviour and nontrivial escape dynamics that can be effectively characterized through survival probabilities and escape statistics \cite{Edelman09}.

Motivated by these developments, we investigate transport and escape in the Stochastic Web Map, an area-preserving dynamical system in which transport is mediated by extended web-like structures in phase space. In contrast to KAM-type systems, transport in the Stochastic Web Map is enabled by global stochastic networks whose geometry is controlled by a discrete symmetry parameter. By introducing absorbing boundaries and analyzing survival probabilities and escape frequencies, we show that in the chaotic regime escape dynamics are governed by a single characteristic time scale, leading to universal behavior that is effectively independent of the underlying symmetry. This provides a clear characterization of transport in a non-KAM system and highlights the role of global structures in controlling escape dynamics.

\section{Model and Scattering Setup}

The Stochastic Web Map (SWM) is a two-dimensional, area-preserving discrete-time dynamical system defined in action-angle variables $(\theta,I)$. Its dynamics arise from the interplay between a nonlinear perturbation and a discrete rotational symmetry, which together generate extended stochastic web structures in phase space that enable global transport. It is given by \cite{Abdullaev07}
\begin{equation}\label{Eq:SWM}
\begin{aligned}
I_{n+1} &= \bigl(I_n + K \sin \theta_n \bigr)\cos\!\left(\frac{2\pi}{q}\right)
          + \theta_n \sin\!\left(\frac{2\pi}{q}\right), \\[6pt]
\theta_{n+1} &= -\bigl(I_n + K \sin \theta_n \bigr)\sin\!\left(\frac{2\pi}{q}\right)
               + \theta_n \cos\!\left(\frac{2\pi}{q}\right).
\end{aligned}
\end{equation}

\begin{figure}[ht]
    \centering
    \includegraphics[width=0.8\linewidth]{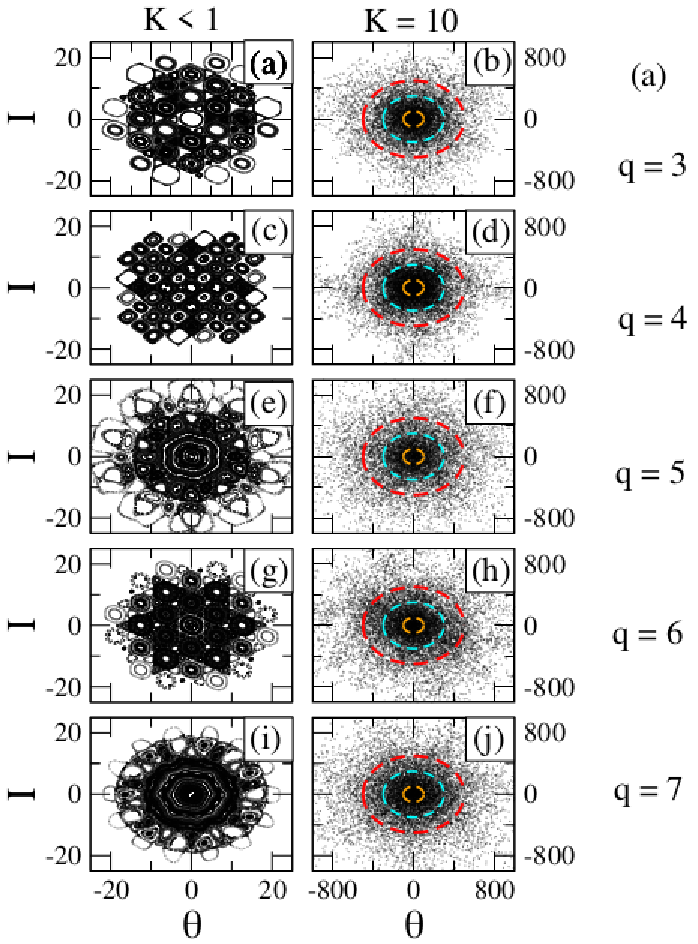}
    \caption{Poincaré surfaces of section for the Stochastic Web Map of Eq. \eqref{Eq:SWM} with $K<1$ (left panels), $K=10$ (right panels), and several values of $q$. A random initial condition with $\theta_{0}\in (-15,15)$ and $I\in(-15,15)$ was iterated $10^4$ times (black dots). The red, cyan and orange dashed lines in panels (b, d, f, h, j) indicate the position of holes in the plane $(\theta,I)$ with radii $h = 500$, $300$ and $100$, respectively.}
    \label{fig:PhaSpaSWM}
\end{figure}

The dynamics is governed by two control parameters: The nonlinearity strength $K$, which controls the strength of chaos, and the integer symmetry parameter $q$, which determines the rotational symmetry of the SWM. As illustrated by the Poincaré surfaces of section in Fig. \ref{fig:PhaSpaSWM}, the map exhibits two distinct diffusive regimes: a slow diffusion regime for small values of $K$ ($K<1$), where transport is constrained along thin web channels, and a quasilinear diffusion regime for large $K$ ($K>10$), where resonance overlap enhances transport. 

In the weakly nonlinear limit, the system remains near-integrable and may formally satisfy the smoothness conditions of KAM theory. However, due to the resonant web structure imposed by the discrete rotational symmetry, invariant tori do not act as effective global transport barriers. Instead of the standard KAM scenario characterized by the progressive breakup of invariant tori, transport is mediated by interconnected web channels that span phase space, allowing diffusion even at low $K$.

To investigate transport and escape, we open the map \eqref{Eq:SWM} by introducing an absorbing boundary defined by a circular hole of radius $h$, centered at the origin, as shown in Fig. \ref{fig:PhaSpaSWM}(right panels). A trajectory is considered to escape once its distance from the origin exceeds the threshold $h$. This choice is consistent with the intrinsic rotational symmetry of the map and avoids directional bias in the escape mechanism. The analysis is performed for hole radii ranging from $h=10$ to $h=900$, allowing us to probe different degrees of openness. 

For each pair ($K$,$h$), we consider an ensemble of $10^{5}$ orbits evolving according to map \eqref{Eq:SWM}. In the quasilinear regime ($K>10$), initial conditions are uniformly distributed within a bounded region of phase space. In contrast, for small values of $K$, initial conditions are selected to avoid trapping inside regular islands and to ensure diffusion along the stochastic web. Once the escape condition is met, the trajectory is removed and a new initial condition is introduced to maintain a constant number of realizations. 

We compute the number of surviving trajectories $N_{s}(n)$ at iteration $n$, from which the survival probability is defined as $P_{\text{S}}(n)=N_{s}(n)/M$, where $M=10^{5}$. By recording the number of trajectories that escape at each iteration, we construct the histogram corresponding to the escape frequency $P_{\text{E}}(n)$ \cite{Mendez23}. We thus analyze the escape dynamics of map \eqref{Eq:SWM} as a function of the parameters $K$ and $h$, which control the nonlinearity and the openness of the system, respectively.

\begin{figure}[htbp]
    \centering
    \includegraphics[width=\linewidth]{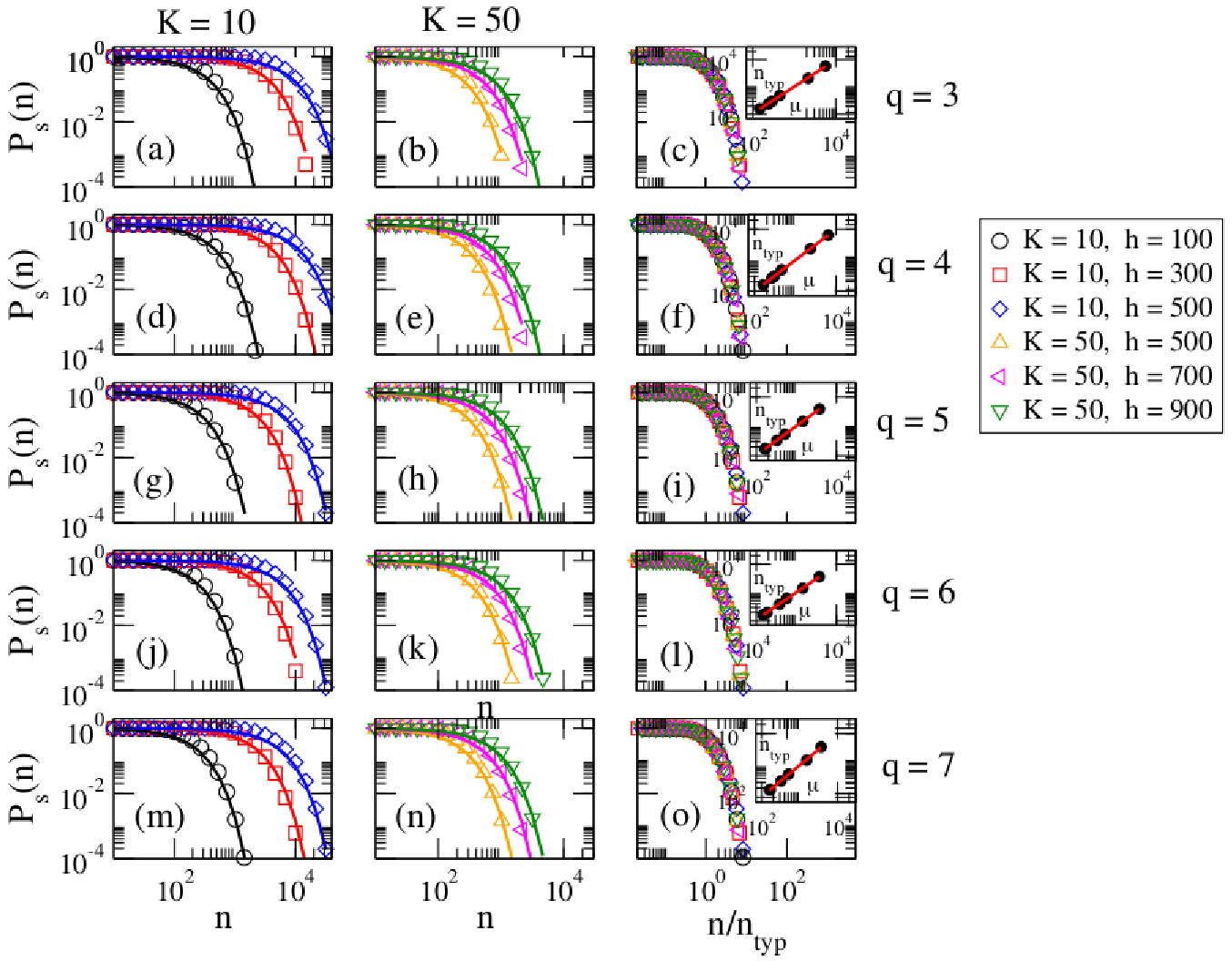}
    \caption{(a, b, d, e, g, h, j, k, m, n) Survival probability $P_{\text{S}}(n)$ as a function of $n$ for the Stochastic Web Map for several combinations of $K$ and $h$, and $q = 3$, $4$, $5$, $6$ and $7$. Full lines correspond to fits of the data using Eq. \eqref{Eq:Ps(n)}, from which $\mu$ is extracted. Curves were computed from an ensemble of $10^7$ trajectories up to $n = 10^{5}$. (c, f, i, l, o). $P_{\text{S}}(n)$ as a function of $n/n_{\text{typ}}$. The relation $n_{\text{typ}} \approx \mu$ (insets) is shown as a reference.}
    \label{fig:SurProbSWM}
\end{figure}

\section{Survival probability and escape frequency}

Escape dynamics provide a powerful framework for characterizing transport processes in open Hamiltonian systems. By coupling a bounded phase space region to an absorbing boundary, transport properties can be quantified through two complementary observables: The survival probability $P_{\text{S}}(n)$ and the escape frequency $P_{\text{E}}(n)$. The former measures the fraction of trajectories that remain inside the interaction region after $n$ iterations, while the latter describes the distribution of escape times. Together, these quantities reveal the characteristic time scales governing diffusion in phase space.

Since the SWM exhibits distinct dynamical behaviors depending on the control parameters, we begin by analyzing the regime in which phase space is effectively ergodic.

\subsection{Ergodic phase space}

To characterize transport in the ergodic regime, we consider relatively large values of the nonlinearity parameter, specifically $K=10$ and $K=50$, which provide reliable statistics within reasonable computational times. For each value of $K$, survival probabilities were computed for several escape radii $h$ and symmetry parameters $q=3$, $4$, $5$, $6$ and $7$. The purpose of varying $q$ is to determine whether the discrete symmetry controlling the geometry of the stochastic web affects the escape dynamics. 

Figure \ref{fig:SurProbSWM} shows $P_{\text{S}}(n)$ as a function of the iteration number $n$. In all cases, $P_{\text{S}}(n)$ starts from unity and decreases monotonically as trajectories diffuse and eventually cross the absorbing boundary. The numerical results reveal that the decay is essentially independent of $q$, indicating that once the system reaches the strongly chaotic regime, transport becomes insensitive to the underlying discrete symmetry. 

The dominant dependence instead arises from $K$ and $h$. Larger values of $K$ enhance chaotic diffusion and therefore lead to faster escape, whereas increasing the radius $h$ enlarges the accessible region and delays escape events. We observe that the survival probability follows an exponential law, 
\begin{equation}\label{Eq:Ps(n)}
    P_{\text{S}}(n) = \exp\left(-\frac{n}{\mu}\right)
\end{equation}
which is a hallmark of strongly chaotic dynamics \cite{Mendez23,Altmann13,Srokowski93}. Continuous curves in Fig. \ref{fig:SurProbSWM} correspond to fits of Eq. \eqref{Eq:Ps(n)} to the numerical data, where $\mu$ is treated as a fitting parameter that characterizes the escape time scale.

\begin{figure}[htbp]
    \centering
    \includegraphics[width=\linewidth]{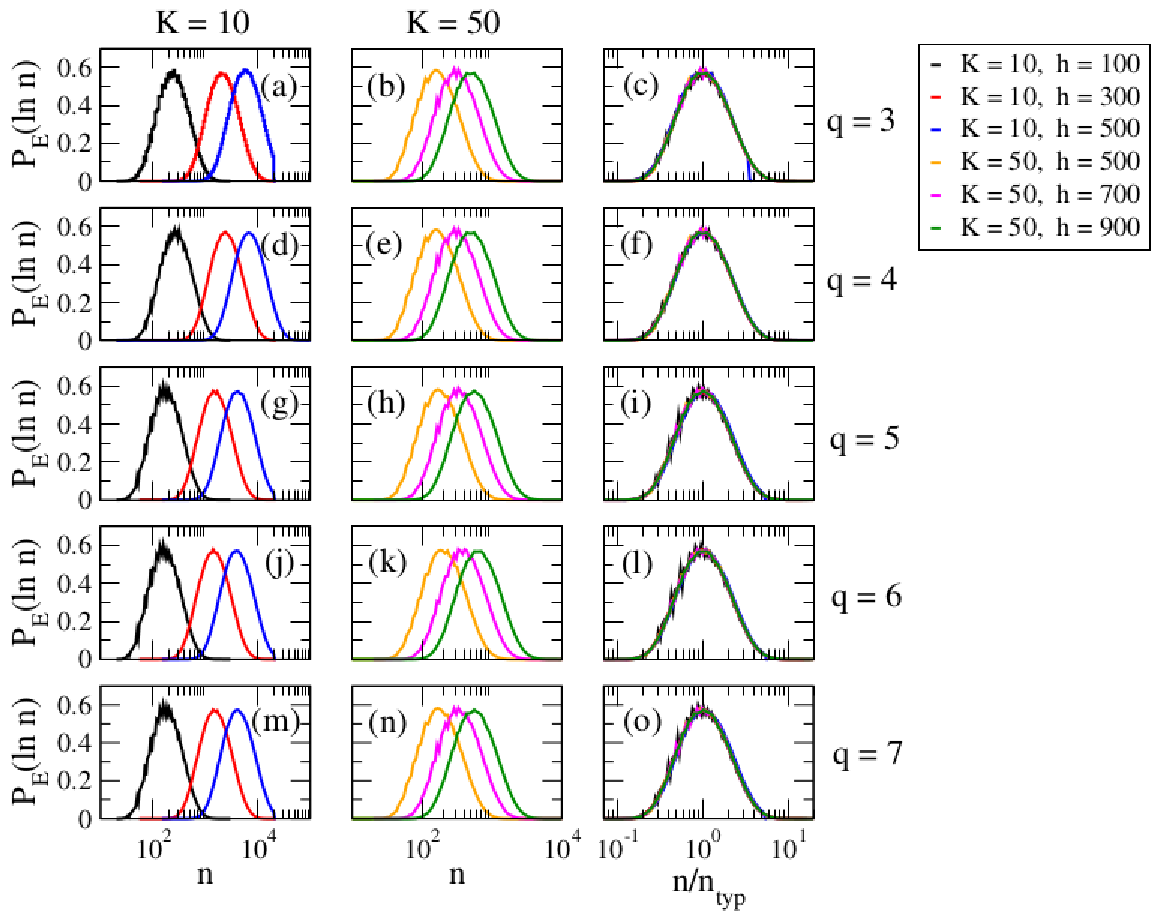}
    \caption{Histograms for the frequency of particle escape $P_{\text{E}}(\text{ln }n)$ when (a, d, g, j, m) $K=10$ and (b, e, h, k, n) $K=50$ for several combinations of $h$, and $q = 3$, $4$, $5$, $6$ and $7$ for the Stochastic Web Map. (c, f, i, l, o) $P_{\text{E}}(\text{ln }n)$ as a function of $n/n_{\text{typ}}$. Each histogram was constructed from an ensemble of $10^{7}$ trajectories.}
    \label{fig:FreEscSWM}
\end{figure}

\subsection{Typical iteration time and scaling behavior}

Previous studies have shown that chaotic escape processes can be described by a characteristic scale known as the typical time \cite{Mendez23}, defined as
\begin{equation}\label{Eq:ntyp}
    n_{typ} = \exp\langle \ln{n}\rangle.
\end{equation}
This quantity provides an accurate estimate of the peak position in the escape-time distribution. The connection between survival probability and escape frequency, $P_{\text{E}}(n) = -dP_{\text{S}}(n)/dn$, implies that the decay constant satisfies $\mu\approx n_{\text{typ}}$ \cite{Mendez23}. Consequently, the survival probability can be rewritten as
\begin{equation}\label{Eq:Ps(n)2}
    P_{\text{S}}(n)\approx\exp\left(-\frac{n}{n_{typ}} \right).
\end{equation}

The numerical observation that $\mu$ and $n_{typ}$ coincide within statistical accuracy is shown in the insets of Fig. \ref{fig:SurProbSWM}. This result has an important implication: both $P_{\text{S}}(n)$ and the escape statistics depend only on the scaled variable $n/n_{\text{typ}}$, making $n_{\text{typ}}$ the relevant time scale for the escape dynamics. Indeed, when plotting $P_{\text{S}}(n)$ as a function of $n/n_{\text{typ}}$, the curves for different combinations of ($K$,$h$,$q$) collapse onto a single universal profile.

\subsection{Escape frequency}

The escape frequency $P_{\text{E}}(n)$ characterizes when trajectories leave the system. Numerically, $P_{\text{E}}(n)$ increases with $n$, reaches a maximum, and then decays exponentially. To better visualize the distribution, we analyze the logarithmic representation $P_{\text{E}}(\ln n)$ \cite{Mendez23}. 

Figure \ref{fig:FreEscSWM} presents $P_{\text{E}}(\ln{n})$ for the same parameter values used in the survival probability analysis. The peak of the distribution coincides with $n_{\text{typ}}$, confirming it as the dominant escape time scale. Furthermore, when the distributions are rescaled by $n/n_{\text{typ}}$, all histograms collapse onto a universal curve, removing the dependence on $h$ and $q$. This demonstrates that, in the ergodic regime, transport is governed by a single characteristic time associated with chaotic diffusion.

While these results share qualitative similarities with those observed in other chaotic systems, such as fractional maps \cite{Mendez23}, the underlying mechanisms are different. In fractional maps, anomalous transport is attributed to memory effects, whereas in the SWM transport is mediated by extended web structures. The observed independence of the escape properties with respect to $q$ indicates that, in the ergodic regime, chaotic diffusion dominates over the geometric details of the web.

\begin{figure}[htbp]
    \centering
    \includegraphics[width=\linewidth]{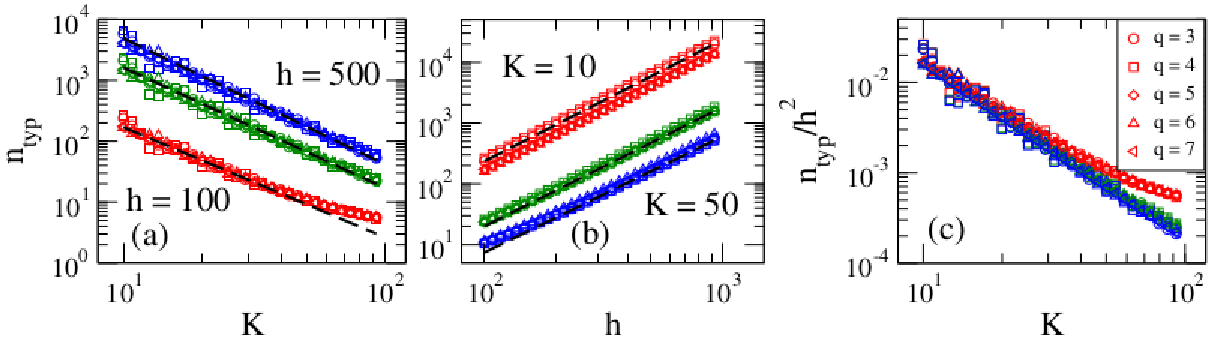}
    \caption{(a) Typical iteration time $n_{\text{typ}}$ as a function of $K$ for the Stochastic Web Map. Here $h =100$ (red symbols), $h=300$ (green symbols), and $h=500$ (blue symbols); different values of $q$ are shown: $q = 3$ ($\circ$), $q = 4$ ($\Box$), $q = 5$ ($\diamond$), $q = 6$ ($\triangle$) and $q = 7$ ($\lhd$). The scaling $n_{\text{typ}}\propto K^{-2}$ (dashed lines) is shown as a reference. (b) $n_{\text{typ}}$ as a function of $h$ for $K=10$ (red symbols), $K=30$ (green symbols), and $K=50$ (blue symbols), for the values of $q$ reported in (a). The scaling $n_{\text{typ}}\propto h^{2}$ (dashed lines) is shown as a reference. (c) Typical escape time $n_{\text{typ}}$ normalized by $h^{2}$, as a function of $K$. Same data as in panel (a).}
    \label{fig:IteTime}
\end{figure}

We further characterize the dependence of $n_{\text{typ}}$ on the system parameters. Figure \ref{fig:IteTime} shows $n_{\text{typ}}$ as a function of $K$ and $h$. Consistent with the previous observations, $n_{\text{typ}}$ exhibits a power-law dependence
\begin{equation}\label{Eq:Fitntyp}
    n_{\text{typ}} \propto K^{\gamma_{K}}h^{\gamma_{h}}.
\end{equation}

Fitting the quasilinear regime yields $\gamma_{K}\approx -2$ and $\gamma_{h}\approx 2$, confirming the scaling $n_{\text{typ}}\propto K^{-2}h^{2}$. However, Fig. \ref{fig:IteTime}(a) and \ref{fig:IteTime}(c) reveal a deviation for $h=100$ and $K>30$. In this region, $n_{\text{typ}}$ remains larger than the $K^{-2}$ prediction, causing the numerical data to lie above the fitting line. This indicates that, for high nonlinearity and small escape regions, the dynamics depart from the standard diffusive expectation, suggesting a breakdown of the quasilinear approximation.

\subsection{Escape in the non-ergodic regime}

\begin{figure}[htbp]
    \centering
    \includegraphics[width=\linewidth]{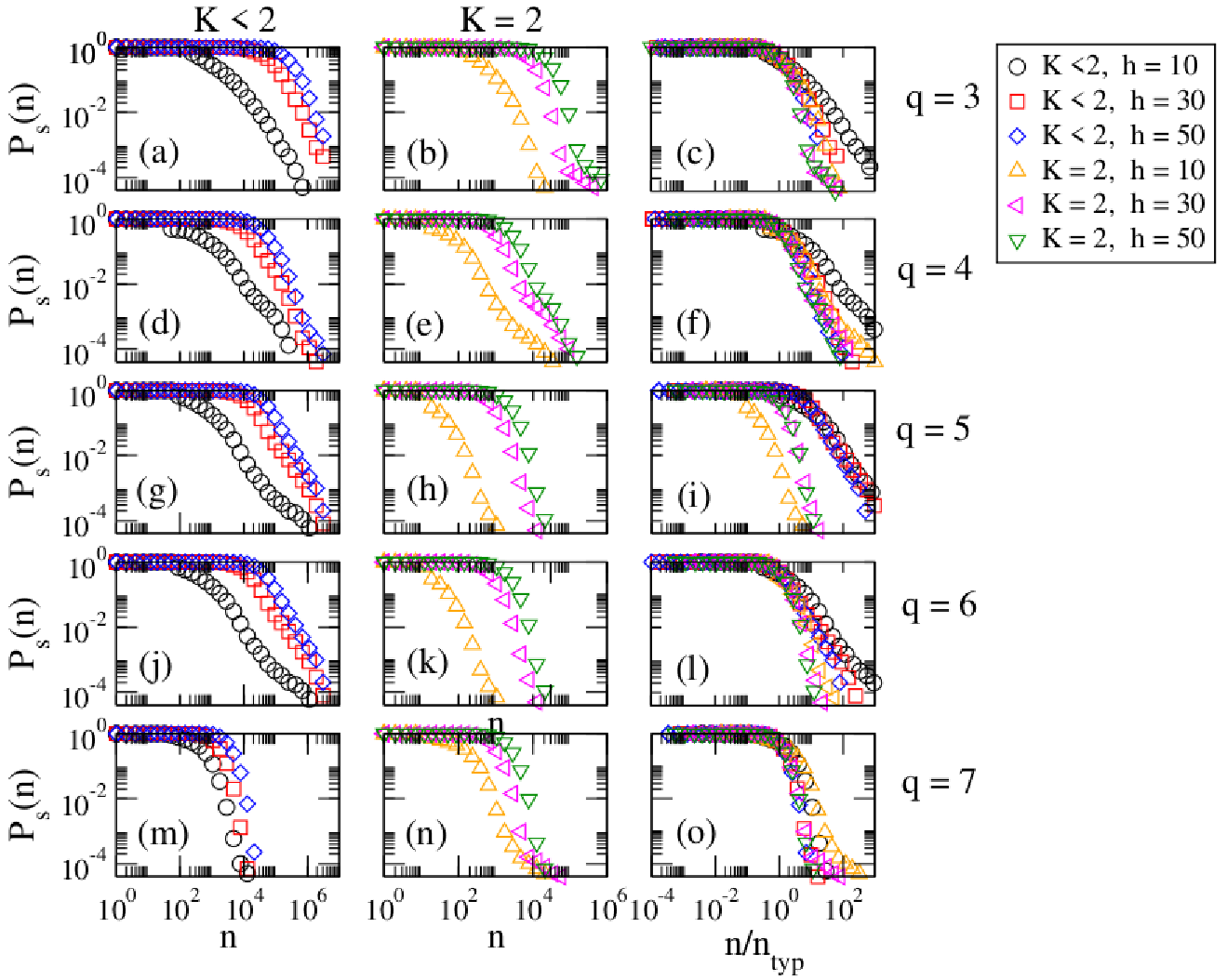}
    \caption{(a, b, d, e, g, h, j, k, m, n) Survival probability $P_{\text{S}}(n)$ as a function of $n$ for the Stochastic Web Map in the non-ergodic regime, with several combinations of $K$ and $h$, and $q = 3$, $4$, $5$, $6$ and $7$. Curves were computed from an ensemble of $10^7$ trajectories up to $n = 10^{5}$. (c, f, i, l, o) $P_{\text{S}}(n)$ as a function of $n/n_{\text{typ}}$. The values of $K$ in (a,d,g,j,m) are ($1.6$, $0.8$, $1.4$, $0.8$, $1.7$), respectively.}
    \label{fig:PsNonErgo}
\end{figure}

In the low-nonlinearity regime ($K\leq 2$), the phase space of the SWM is characterized by a mixed structure of stable islands and narrow chaotic channels. In this case, global transport is governed by slow diffusion, and the escape dynamics deviate significantly from the universality observed in the ergodic regime. 

As shown in Fig. \ref{fig:PsNonErgo}, $P_{\text{S}}(n)$ no longer follows a simple exponential decay, exhibiting instead a complex multi-stage behavior due to the stickiness of trajectories near stable structures. In this regime, $\mu$ no longer coincides with $n_{\text{typ}}$, and the scaling law breaks down. This indicates that escape is controlled by the local topology of the mixed phase space rather than by a single global time scale.

\begin{figure}[htbp]
    \centering
    \includegraphics[width=\linewidth]{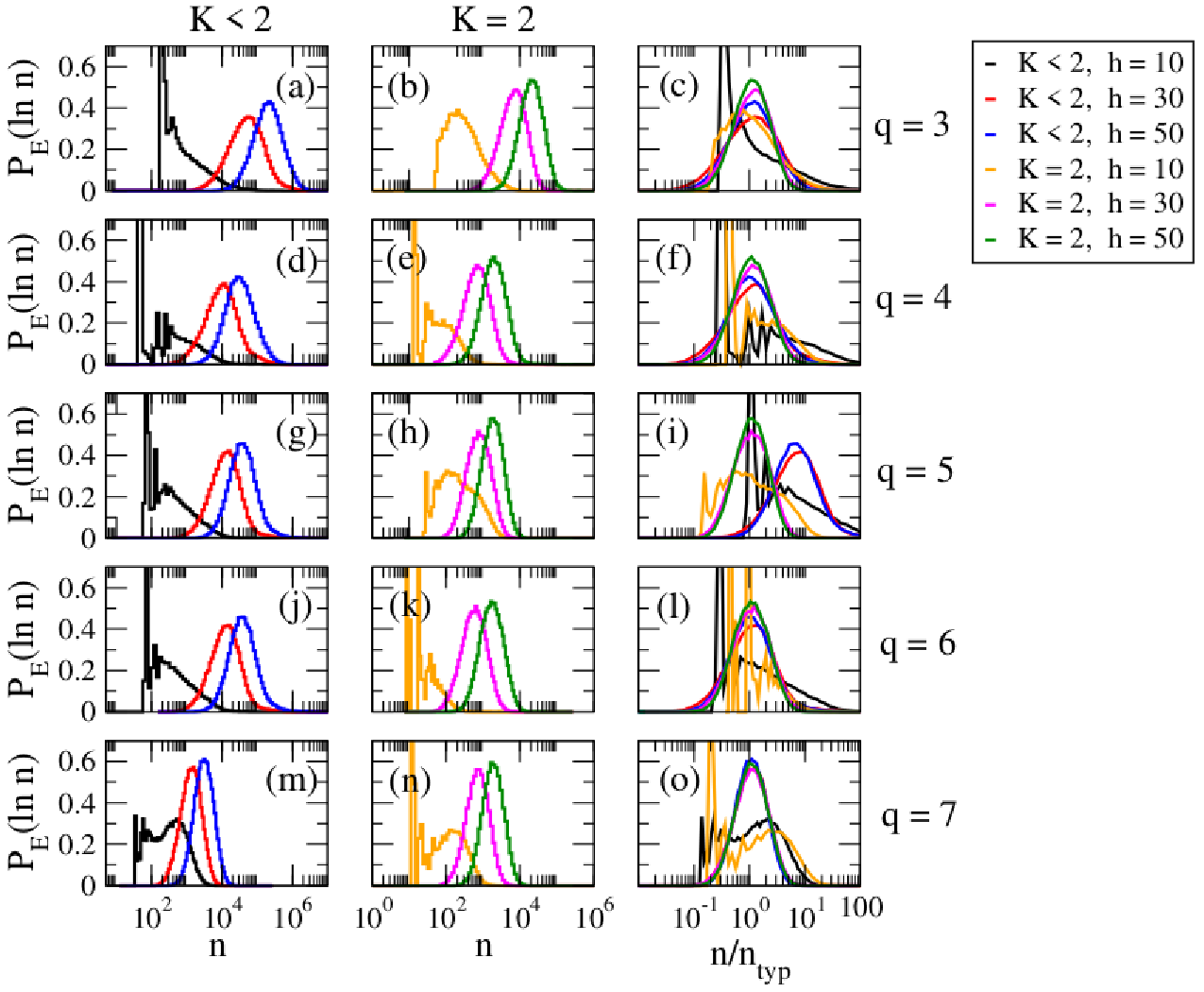}
    \caption{Histograms for the escape frequency $P_{\text{E}}(\text{ln }n)$ in the non-ergodic regime when (a, d, g, j, m) $K<2$ and (b, e, h, k, n) $K=2$ for several combinations of $h$, and $q = 3$, $4$, $5$, $6$ and $7$, for the Stochastic Web Map. (c, f, i, l, o) $P_{\text{E}}(\text{ln }n)$ as a function of $n/n_{\text{typ}}$. Each histogram was constructed from an ensemble of $10^{7}$ trajectories. The values of $K$ for (a,d,g,j,m) are ($1.6$, $0.8$, $1.4$, $0.8$, $1.7$), respectively.}
    \label{fig:PENonErgo}
\end{figure}

Unlike the chaotic regime, the symmetry parameter $q$ explicitly influences the escape rate. We find that escape is faster for $q=7$ than for lower symmetry orders. Since initial conditions were restricted to lie within the stochastic web, this dependence reflects how the connectivity of the web dictates transport pathways toward the hole. Finally, $P_{\text{E}}(\ln n)$ is highly sensitive to $h$, as shown in Fig. \ref{fig:PENonErgo}, where the smallest radius produces irregular histograms with pronounced early peaks. This confirms that, in the non-ergodic SWM, transport properties are non-universal and strongly modulated by the resonant web configuration.

\section{Conclusions}

In this work, we have characterized the transport and escape properties of the Stochastic Web Map (SWM) by introducing circular absorbing boundaries of radius $h$. Our results reveal two distinct dynamical regimes, determined by the nonlinearity strength $K$ and the underlying phase space structure. In the ergodic regime ($K\geq 10$), both the survival probability $P_{\text{S}}(n)$ and the escape frequency $P_{\text{E}}(n)$ exhibit universal scaling governed by a single characteristic time $n_{\text{typ}}$. In this regime, escape dynamics are effectively independent of the symmetry parameter $q$, indicating that global chaotic diffusion dominates over the geometric details of the stochastic web. We find that $n_{\text{typ}}\propto K^{-2}h^{2}$, in agreement with quasilinear diffusion theory. However, for small hole sizes and large nonlinearity (e.g., $h=100$ and $K>30$), deviations from this scaling are observed, signaling a breakdown of the quasilinear approximation and indicating that transport cannot be fully described within a simple diffusive framework.

In the non-ergodic regime ($K<10$), the universal behavior is lost. The presence of stable islands and the stickiness of trajectories near invariant structures lead to a multi-stage decay of $P_{\text{S}}(n)$ and to a breakdown of the scaling with $n_{\text{typ}}$. In contrast to the ergodic case, escape dynamics become sensitive to the symmetry parameter $q$, reflecting the role of the web connectivity in shaping transport pathways. The absence of collapse in the rescaled escape distributions further confirms that transport in this regime is non-universal and strongly influenced by the local phase space topology. Overall, our results establish $n_{\text{typ}}$ as the fundamental time scale controlling escape dynamics in the chaotic regime and demonstrate that transport is governed by global phase-space structures rather than by symmetry. These findings provide a unified characterization of escape in a non-KAM system and open new perspectives for understanding transport in systems with extended stochastic networks.

\section*{Ackowledgments}
E.D.L. acknowledges support from Brazilian agencies CNPq (No.~301318/2019-0, 304398/2023-3) and FAPESP (No.~2019/14038-6 and No.~2021/09519-5). J.A.M.-B. thanks support from SECIHTI (Grant No.~CBF-2025-I-2236), Mexico.


\end{document}